\documentclass[11pt,leqno,twoside]{article}

\usepackage{amsfonts,amsmath,amsthm,amssymb}
\usepackage{enumerate}
\usepackage{graphics,graphicx,subfigure}

\usepackage{color}
\usepackage{todonotes}
\usepackage{cancel}
\usepackage{url}
\usepackage{hyperref}
\usepackage{makeidx}
\usepackage{showidx}
\usepackage{multicol}        
\usepackage{xspace}
\usepackage{stmaryrd}        
\usepackage{pifont}          
\usepackage{fancybox}        
 \usepackage{fancyhdr}
\theoremstyle{plain}
 \theoremstyle{definition}
 \newtheorem{lem}{Lemma}
 \newtheorem{defn}[lem]{Definition}
 \newtheorem{thm}[lem]{Theorem}
 \newtheorem{prop}[lem]{Proposition}
 \newtheorem{cor}[lem]{Corollary}
 \newtheorem{notn}[lem]{Notations}
 \newtheorem{pb}[lem]{Problem}
 \newtheorem{form}[lem]{Formulae}
 \newtheorem{rk}{Remark}
 
 \newtheorem*{com}{Comment}
 \newtheorem*{ex}{Example}
 \theoremstyle{remark}

 \newcommand{\blem}{\begin{lem}}
 \newcommand{\elem}{\end{lem}}
 \newcommand{\bdefn}{\begin{defn}}
 \newcommand{\edefn}{\end{defn}}
 \newcommand{\bthm}{\begin{thm} }
 \newcommand{\ethm}{\end{thm}}
 \newcommand{\bprop}{\begin{prop}}
 \newcommand{\eprop}{\end{prop}}
 \newcommand{\bcor}{\begin{cor}}
 \newcommand{\ecor}{\end{cor}}
 \newcommand{\bnotn}{\begin{notn}}
 \newcommand{\enotn}{\end{notn}}
 \newcommand{\bpb}{\begin{pb}}
 \newcommand{\epb}{\end{pb}}
 \newcommand{\bform}{\begin{form}}
 \newcommand{\eform}{\end{form}}
 \newcommand{\brk}{\begin{rk}}
 \newcommand{\erk}{\end{rk}}
 \newcommand{\bcom}{\begin{com}}
 \newcommand{\ecom}{\end{com}}
 \newcommand{\bex}{\begin{ex}}
 \newcommand{\eex}{\end{ex}}
 \newcommand{\bpf}{\begin{proof}}
 \newcommand{\epf}{\end{proof}}

\newcommand{\va}{{\bf a}}

\newcommand{\ve}{{\bf e}}
\newcommand{\vf}{{\bf f}}
\newcommand{\vg}{{\bf g}}
\newcommand{\vh}{{\bf h}}

\newcommand{\vu}{{\bf u}}
\newcommand{\vv}{{\bf v}}

\newcommand{\vx}{{\bf x}}
\newcommand{\vy}{{\bf y}}
\newcommand{\vz}{{\bf z}}
\newcommand{\vA}{{\bf A}}

\newcommand{\vD}{{\bf D}}

\newcommand{\vI}{{\bf I}}

\newcommand{\vM}{{\bf M}}

\newcommand{\cA}{\mathcal{A}}

\newcommand{\cK}{\mathcal{K}}

\newcommand{\cN}{\mathcal{N}}

\newcommand{\cP}{\mathcal{P}}

\newcommand{\bE}{\mathbb{E}}

\newcommand{\bP}{\mathbb{P}}

\newcommand{\bR}{\mathbb{R}}

\newcommand{\be}{\begin{equation}}
\newcommand{\ee}{\end{equation}}
\newcommand{\bal}{\begin{align}}
\newcommand{\eal}{\end{align}}
\newcommand{\ba}{\begin{align*}}
\newcommand{\ea}{\end{align*}}
\newcommand{\bmx}{\begin{matrix}}
\newcommand{\emx}{\end{matrix}}
\newcommand{\bbmx}{\begin{bmatrix}}
\newcommand{\ebmx}{\end{bmatrix}}
\newcommand{\bpmx}{\begin{pmatrix}}
\newcommand{\epmx}{\end{pmatrix}}
\newcommand{\bvmx}{\begin{vmatrix}}
\newcommand{\evmx}{\end{vmatrix}}

\newcommand{\ol}{\overline}
\newcommand{\wh}{\widehat}
\newcommand{\wt}{\widetilde}
\newcommand{\f}{\frac}
\newcommand{\df}{\dfrac}

\newcommand{\imp}{\Longrightarrow}

\newcommand{\inc}{\subseteq}

\newcommand{\ip}[2]{\langle {#1}, {#2} \rangle}

\newcommand{\sgn}{\mathrm{sgn}}
\newcommand{\supp}{\mathrm{supp}}

\newcommand{\argmin}{{\rm argmin}\,}

\newcommand{\minimize}[1]{\underset{#1}{\rm minimize}\,}

\newcommand{\la}{\lambda}

\newcommand{\eps}{\varepsilon}

\title{\vspace{-20mm}One-Bit Compressive Sensing of Dictionary-Sparse Signals\medskip\hrule height 1.2pt \vspace{-6mm}}
\author{R. Baraniuk, S. Foucart, D. Needell, Y. Plan, and M. Wootters}
\date{\vspace{-6mm}\rule{100mm}{0.8pt}}

\newcommand\shorttitle{One-Bit Compressive Sensing of Dictionary-Sparse Signals}
\newcommand\authors{R. Baraniuk, S. Foucart, D. Needell, Y. Plan, M. Wootters}

\begin{document}
\maketitle

\vspace{-15mm}
\begin{abstract}
One-bit compressive sensing has extended the scope of sparse recovery by showing that sparse signals can be accurately reconstructed even when their linear measurements are subject to the extreme quantization scenario of binary samples---only the sign of each linear measurement is maintained.  
Existing results in one-bit compressive sensing rely on the assumption that the signals of interest are sparse in some fixed orthonormal basis.  
However, in most practical applications, signals are sparse with respect to an overcomplete dictionary, rather than a basis.
There has already been a surge of activity to obtain recovery guarantees under such a generalized sparsity model in the classical compressive sensing setting.  
Here, we extend the one-bit framework to this important model, providing a unified theory of one-bit compressive sensing under dictionary sparsity.   
Specifically, we analyze several different algorithms---based on convex programming and on hard thresholding---and show that, under natural assumptions on the sensing matrix (satisfied by Gaussian matrices), these algorithms can efficiently recover analysis-dictionary-sparse signals in the one-bit model.
\end{abstract}

\noindent {\it Key words and phrases:}  compressive sensing, quantization, one-bit compressive sensing, tight frames, convex optimization, thresholding.

\vspace{-5mm}
\begin{center}
\rule{100mm}{0.8pt}
\end{center}


\section{Introduction}\label{sec:intro}

The basic insight of compressive sensing is that a small number of linear measurements can be used to reconstruct sparse signals.  
In traditional compressive sensing, we wish to reconstruct an $s$-sparse\footnote{A signal $\vx \in \bR^N$ is called $s$-sparse if $\|\vx\|_0 := |\supp(\vx)| \leq s \ll N$.} signal $\vx \in \bR^N$ from linear measurements of the form
\begin{equation}\label{meas}
\vy = \vA\vx \in \bR^m \qquad\text{(or its corrupted version $\vy = \vA\vx + \ve$)}, 
\end{equation}
where $\vA$ is a $m\times N$ measurement matrix.  A significant body of work over the past decade has demonstrated that the $s$-sparse (or nearly $s$-sparse) signal $\vx$ can be accurately and efficiently recovered from its measurement vector $\vy = \vA\vx$ when $\vA$ has independent Gaussian entries, say, and when $m \asymp s\log(N/s)$ \cite{DSPweb,eldar2012compressed,foucart2013}.  

This basic model has been extended in several directions.  
Two important ones---which we focus on in this work---are (a) extending the set of signals to include the larger and important class of \em dictionary sparse \em signals and (b) considering highly quantized measurements as in \em one-bit compressive sensing. \em 

Both of these settings have important practical applications and have received much attention in the past few years.  However, to the best of our knowledge, they have not been considered together before.  In this work, we extend the theory of one-bit compressive sensing to dictionary sparse signals. 
Below, we briefly review the background on these notions, set up notation, and outline our contributions.

\subsection{One-bit measurements}
\label{ssec:onebitintro}

In practice, each entry $y_i = \langle \va_i, \vx\rangle$ (where $\va_i$ denotes the $i$th row of $\vA$) of the measurement vector in \eqref{meas} needs to be quantized.  
That is, rather than observing $\vy=\vA\vx$,
one observes $\vy = Q(\vA\vx)$ instead,
where $Q: \bR^m \rightarrow \cA$ denotes the quantizer that maps each entry of its input to a corresponding quantized value in an alphabet $\cA$.  
The so-called \textit{one-bit compressive sensing} \cite{Boufounos2008} problem refers to the case when $|\cA| = 2$ and one wishes to recover $\vx$ from its heavily quantized (one bit) measurements $\vy = Q(\vA\vx)$.  
The simplest quantizer in the one-bit case uses the alphabet $\cA = \{-1, 1\}$ and acts by taking the sign of each component as
\begin{equation}\label{eq:quantized}
y_i = Q(\langle \va_i, \vx\rangle) = \sgn(\langle \va_i, \vx\rangle),
\end{equation}
which we denote in shorthand by $\vy = \sgn(\vA\vx)$.  
Since the publication of \cite{Boufounos2008} in 2008, 
several efficient methods, both iterative and optimization-based, 
have been developed to recover the signal~$\vx$ (up to normalization) from its one-bit measurements (see e.g. \cite{pv-1-bit,pv-noisy-1bit,gnjn2013,Jacques2011,yan2012robust,jacques2013quantized}).  
In particular, it is shown \cite{Jacques2011} that the direction of any $s$-sparse signal $\vx$ can be estimated by some $\hat{\vx}$ produced from $\vy$ with accuracy 
$$
\left\| \frac{\vx}{\|\vx\|_2} - \frac{\hat{\vx}}{\|\hat{\vx}\|_2}\right\|_2 \leq \eps 
$$
when the number of measurements is at least 
$$m = \Omega\left( \frac{ s \ln( N/s ) }{\eps} \right).$$
Notice that with measurements of this form, we can only hope to recover the direction of the signal, not the magnitude.  However, we can recover the entire signal if we allow for \em thresholded measurements \em of the form
\begin{equation}\label{eq:quantizeddither}
y_i = \sgn( \ip{\va_i}{\vx} - \tau_i ).
\end{equation}
In practice, it is often feasible to obtain quantized measurements of this form, and they have been studied before. 
Existing works using measurements of the form~\eqref{eq:quantizeddither} have also allowed for \em adaptive \em thresholds; that is, the $\tau_i$ can be chosen adaptively based on $y_j$ for $j < i$.  The goal of those works was to improve the convergence rate, i.e., the dependence on $\eps$ in the number of measurements $m$.  
It is known that a dependence of $\Omega(1/\eps)$ is necessary with nonadaptive measurements, but recent work on Sigma-Delta quantization \cite{saab2015quantization} and other schemes
\cite{exponentialBFNPW14,knudson2014one} have shown how to 
break this barrier using measurements of the form~\eqref{eq:quantizeddither} with adaptive thresholds. 

In this article, we do not focus on the decay rate (the dependence on $\eps$), nor do we consider adaptive measurements.  However, we do consider nonadaptive measurements both of the form \eqref{eq:quantized} and~\eqref{eq:quantizeddither}.  This allows us to provide results on reconstruction of the magnitude of signals, as well as the direction.

\subsection{Dictionary Sparsity}
\label{ssec:dicsps}
Although the classical setting assumes that the signal $\vx$ itself is sparse, 
most signals of interest are not immediately sparse.  In the straightforward case, a signal may be instead sparse after some transform;
for example, images are known to be sparse in the wavelet domain, sinusoidal signals in the Fourier domain, and so on \cite{Daube_Ten}.  Fortunately, the classical framework extends directly to this model, since the product of a Gaussian matrix and an orthonormal basis is still Gaussian.

However, in many practical applications the situation is not so straightforward, and the signals of interest are sparse not in an orthonormal basis but rather in a redundant (highly overcomplete) dictionary; this is known as \em dictionary sparsity. \em
Signals in radar and sonar systems, for example, are sparsely represented in Gabor frames, which are highly overcomplete and far from orthonormal \cite{FS98:Gabor-Analysis}.  
Images may be sparsely represented in curvelet frames \cite{CDDY05:Fast,CD02:New-Tight}, undecimated wavelet frames \cite{SED04:Redundant}, and other frames which by design are highly redundant.  Such redundancy allows for sparser representations and a wider class of signal representations.  Even in the Fourier domain, utilizing an oversampled DFT allows for much more realistic and practical signals to be represented.  For these reasons, recent research has extended the compressive sensing framework to the setting where the signals of interest are sparsified by overcomplete tight frames (see e.g. \cite{rauhut2008compressed,RefWorks:60,RefWorks:607,Foucart2016}).   

Throughout this article, we consider a dictionary $\vD \in \bR^{n \times N}$
which is assumed to be a tight frame,
in the sense that 
$$
\vD \vD^* = \vI_n.
$$  
To distinguish between the signal and its sparse representation, we write $\vf\in\bR^n$ for the signal of interest and $\vf=\vD\vx$, where $\vx\in\bR^N$ is a sparse coefficient vector.  We then acquire the samples of the form $\vy = \vA\vf = \vA\vD\vx$ and attempt to recover the signal $\vf$.  
Note that, due to the redundancy of~$\vD$,
we do not hope to be able to recover a unique coefficient vector $\vx$.  In other words, even when the measurement matrix $\vA$ is well suited for sparse recovery, the product $\vA\vD$ may have highly correlated columns,
making recovery of $\vx$ impossible.  With the introduction of a noninvertible sparsifying transform $\vD$, it becomes important to distinguish between two related but distinct notions of sparsity. Precisely, we say that \vspace{-5mm}
\begin{itemize}
\item $\vf$ is $s$-synthesis-sparse if $\vf = \vD \vx$ for some $s$-sparse $\vx \in \bR^N$;
\item $\vf$ is $s$-analysis-sparse if $\vD^* \vf \in \bR^N$ is $s$-sparse.
\end{itemize}\vspace{-5mm}
We note that analysis sparsity is a stronger assumption because, assuming analysis sparsity, one can always take $\vx = \vD^* \vf$ in the synthesis sparsity model.  See  \cite{elad2007analysis} for an introduction to the analysis sparse model in compressive sensing (also called the analysis cosparse model). 

Instead of exact sparsity, it is often more realistic to study \em effective sparsity. \em 
We call a coefficient vector $\vx \in \bR^N$ effectively $s$-sparse if
$$
\|\vx\|_1 \le \sqrt{s} \|\vx\|_2,
$$ 
and we say that \vspace{-5mm}
\begin{itemize}
\item $\vf$ is effectively $s$-synthesis-sparse if $\vf = \vD \vx$ for some effectively $s$-sparse $\vx \in \bR^N$;
\item $\vf$ is effectively $s$-analysis-sparse if $\vD^* \vf \in \bR^N$ is effectively $s$-sparse.
\end{itemize}

We use the notation
\begin{align*}
\Sigma^N_s
& \mbox{ for the set of $s$-sparse coefficient vectors in $\bR^N$,} \\
\Sigma_s^{N,{\rm eff}}
& \mbox{ for the set of effectively $s$-sparse coefficient vectors in $\bR^N$.}
\end{align*}
We also use the notation $B_2^n$ for the set of signals with $\ell_2$-norm at most $1$ (i.e., the unit ball in $\ell_2^n$)
and $S^{n-1}$ for the set of signals with $\ell_2$-norm equal to $1$
(i.e., the unit sphere in $\ell_2^n$).

It is now well known that, if $\vD$ is a tight frame and $\vA$ satisfies analogous conditions to those in the classical setting (e.g., has independent Gaussian entries), then a signal $\vf$ which is (effectively) analysis- or synthesis-sparse can be accurately recovered from traditional compressive sensing measurements $\vy = \vA \vf = \vA\vD\vx$ (see e.g. \cite{rauhut2008compressed,RefWorks:12,RefWorks:60,Paper5,RefWorks:607,RefWorks:581,peleg2013performance,Foucart2016}).

\subsection{One-bit measurements with dictionaries: our setup}

In this article, we study one-bit compressive sensing for dictionary-sparse signals. 
Precisely, our aim is to recover signals $\vf \in \bR^n$ from the binary measurements
$$
y_i = \sgn \langle \va_i, \vf \rangle ,
\qquad i=1,\ldots,m,
$$
or
$$
y_i = \sgn \left( \langle \va_i, \vf \rangle - \tau_i \right), \qquad i = 1,\ldots,m,
$$
when these signals are sparse with respect to a dictionary $\vD$.

As in Section~\ref{ssec:dicsps}, there are several ways to model signals which are sparse with respect to $\vD$. 
In this work, two different signal classes are considered.
For the first one, which is more general, our results are based on convex programming.  
For the second one, which  is a more restrictive, we can obtain results using a computationally simpler algorithm based on hard thresholding.

The first class consists of signals $\vf \in (\vD^*)^{-1} \Sigma_s^{N,\rm{eff}}$ that are effectively $s$-analysis-sparse, i.e., they satisfty
\be 
\label{Assumption}
\|\vD^* \vf\|_1 \le  \sqrt{ s} \|\vD^* \vf\|_2.
\ee
This occurs, of course, when $\vD^* \vf$ is genuinely sparse (analysis sparsity)
and this is realistic if we are working e.g. with piecewise-constant images, since they are sparse after application of the total variation operator.
We consider effectively sparse signals since genuine analysis sparsity is unrealistic when $\vD$ has columns in general position, as it would imply that $\vf$ is orthogonal to too many columns of $\vD$.

The second class consists of signals 
$\vf \in \vD(\Sigma_s^N) \cap (\vD^*)^{-1} \Sigma_{\kappa s}^{N, \rm{eff}}$ 
that are both
$s$-synthesis-sparse and $\kappa s$-analysis-sparse for some $\kappa \ge 1$.
This will occur as soon as the signals are $s$-synthesis-sparse,
 provided we utilize suitable dictionaries $\vD \in \bR^{n \times N}$.
One could take, for instance,
the matrix of an equiangular tight frame when $N = n + k$, $k = {\rm constant}$.
Other examples of suitable dictionaries found in \cite{krahmer2015compressive} include harmonic frames again with $N = n + k$, $k = {\rm constant}$, as well as Fourier and Haar frames with constant redundancy factor $N/n$.

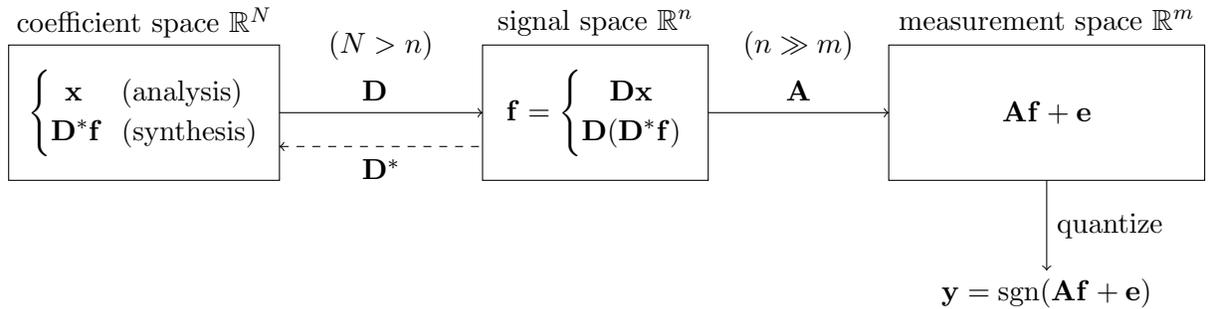
\begin{figure}[ht]
\begin{center}
\begin{tikzpicture}[scale=3]
\node at (0.5,0.4) {coefficient space $\bR^N$};
\draw (-0.1,-0.3) rectangle (1.1,0.3);
\node at (1.55,0.3) {$(N>n)$};
\node at (1.55,0.1) {$\vD{\phantom{^*}}$};
\draw[->] (1.1,0.0) -- (2,0.0);
\draw[dashed, <-] (1.1,-0.15) -- (2,-0.15);
\node at (1.55,-0.25) {$\vD^*$};
\node at (0.5,0) {$\left\{ \bmx \vx \text{\ \ \ \ (analysis)} \\ \vD^* \vf \text{\ \ (synthesis)} \emx\right.$};
\node at (2.5,0.4) {signal space $\bR^n$};
\draw (2,-0.3) rectangle (3,0.3);
\node at (2.5,0) {$\vf = \left\{ \bmx \vD \vx\\ \vD(\vD^* \vf)\emx \right.$};
\draw[->] (3,0) -- (3.8,0);
\node at (3.4,0.3) {$(n \gg m)$};
\node at (3.4,0.1) {$\vA$};
\node at (4.5,0.4) {measurement space $\bR^m$};
\draw (3.8,-0.3) rectangle (5.2,0.3);
\node at (4.5,0) {$\vA \vf + \ve$};
\draw[->] (4.5,-0.3) -- (4.5,-0.7);
\node[anchor=west] at (4.5,-0.5) {quantize};
\node at (4.5,-0.8) {$\vy = \sgn(\vA \vf + \ve)$};
\end{tikzpicture}
\end{center}
\caption{The coefficient, signal, and measurement domains.}
\label{FigSituation}
\end{figure}

Figure \ref{FigSituation} summarizes the relationship between the various domains we deal with.

\subsection{Contributions}

Our main results demonstrate that one-bit compressive sensing is viable even when the sparsifying transform is an overcomplete dictionary. 
As outlined in Section~\ref{ssec:onebitintro}, we consider both the challenge of recovering the direction $\vf/\|\vf\|_2$ of a signal $\vf$, as well as the challenge of recovering the entire signal (direction and magnitude). 
Using measurements of the form $y_i = \sgn\langle \va_i, \vf \rangle$,
we can recover the direction  but not the magnitude;  
using measurements of the form   
 $y_i = \sgn\left( \langle \va_i, \vf \rangle - \tau_i \right)$,
 we may recover both.

In (one-bit) compressive sensing, two standard families of algorithms are (a) algorithms based on convex programming, and (b) algorithms based on thresholding.
In this article, we analyze algorithms from both classes.  One reason to study multiple algorithms is to give a more complete landscape of this problem.  Another reason is that the different algorithms come with different trade-offs (between computational complexity and the strength of assumptions required), and it is valuable to explore this space of trade-offs.

\paragraph{Recovering the direction.}
First, we show that the direction of a dictionary sparse signal can be estimated from one-bit measurements of the type $\sgn(\vA \vf )$.
We consider two algorithms; our first approach is based on linear programming, and our second is based on hard thresholding.  The linear programming approach is more computationally demanding, but applies to a broader class of signals.

In Section~\ref{SecDir}, we prove that both of these approaches are effective, provided the sensing matrix $\vA$ satisfies certain properties.  In Section~\ref{SecTec}, we state that these properties are in fact satisfied by a matrix $\vA$ populated with independent Gaussian entries.  We combine all of these results to prove the statement below.
As noted above, the different algorithms require different definitions of ``dictionary sparsity".
In what follows, $\gamma, C, c$ refer to absolute numerical constants.

\bthm[Informal statement of direction recovery]\label{informal1}
Let $\eps > 0$,
let $m \ge C \eps^{-7} s \ln(eN/s)$,
and let $\vA \in \bR^{m \times n}$ be populated by independent standard normal random variables.
Then, with failure probability at most $\gamma \exp(-c \eps^2 m)$,
any dictionary sparse%
\footnote{\label{note:dicsps}Here, 
``dictionary sparsity" means effective $s$-analysis-sparsity if $\wh{\vf}$ is produced by convex programming
and genuine $s$-synthesis sparsity together with effective $\kappa s$-analysis-sparsity
if $\wh{\vf}$ is produced by hard thresholding. }
 signal $\vf \in \bR^n$ observed via $\vy = \sgn(\vA \vf)$
can be approximated by the output $\wh{\vf}$ of an efficient algorithm with error
$$
\left\| \f{\vf}{\|\vf\|_2} - \f{\wh{\vf}}{\|\wh{\vf}\|_2} \right\|_2 \le \eps.
$$
\ethm
\hspace{0mm}

\paragraph{Recovering the whole signal.}
By using one-bit measurements of the form $\sgn( \vA \vf - \boldsymbol{\tau})$, 
where $\tau_1,\ldots,\tau_m$ are properly normalized Gaussian random thresholds, 
we are able to recover not just the direction but also the magnitude of a dictionary-sparse signal $\vf$.

We consider three algorithms; our first approach is based on linear programming, 
our second approach on second-order cone programming, 
and our third approach on hard thresholding.  Again, there are different trade-offs to the different algorithms.  As above, the approach based on hard thresholding is more efficient, while the approaches based on convex programming apply to a broader signal class.  There is also a trade-off between linear programming and second-order cone programming: the second-order cone program requires knowledge of $\|\vf\|_2$ while the linear program does not (although it does require a loose bound), but the second-order cone programming approach applies to a slightly larger class of signals.

We show in Section~\ref{SecDirMag} that all three of these algorithms are effective when the sensing matrix $\vA$ is populated with independent Gaussian entries and when the thresholds $\tau_i$ are also independent Gaussian random variables.
We combine the results of Section~\ref{SecDirMag} in the following theorem. 

\bthm[Informal statement of signal estimation]
Let $\eps, r, \sigma > 0$,
let $m \ge C \eps^{-9} s \ln(eN/s)$, 
and let $\vA \in \bR^{m \times n}$ and $\boldsymbol{\tau} \in \bR^m$ be populated by independent mean-zero normal random variables with variance $1$ and $\sigma^2$, respectively.  
Then, with failure probability at most $\gamma \exp(-c \eps^2 m)$,
any dictionary sparse\footnote{See footnote \ref{note:dicsps}.} signal $\vf \in \bR^n$ with $\|\vf\|_2 \le r$ observed via $\vy = \sgn(\vA \vf - \boldsymbol{\tau})$
is approximated by the output $\wh{\vf}$ of an efficient algorithm with error
$$
\left\| \vf - \wh{\vf} \right\|_2 \le \eps r.
$$
\ethm

We have not spelled out the dependence of the number of measurements and the failure probability on the parameters $r$ and $\sigma$:
as long as they are roughly the same order of magnitude,
the dependence is absorbed in the constants $C$ and $c$
 (see Section~\ref{SecDirMag} for precise statements).  
As outlined earlier, an estimate of $r$ is required to implement the second-order cone program, but the other two algorithms do not require such an estimate.

\subsection{Discussion and future directions}

The purpose of this work is to demonstrate that techniques from one-bit compressive sensing can be effective for the recovery of dictionary-sparse signals
and we propose several algorithms to accomplish this for various notions of dictionary sparsity.  
Still, some interesting future directions remain.

Firstly, we do not believe that the dependence on $\eps$ above is optimal. 
We do believe instead that a logarithmic dependence on $\eps$ for the number of measurements (or equivalently an exponential decay in the oversampling factor $\la = m / (s \ln(eN/s))$ for the recovery error )
is possible by choosing the thresholds $\tau_1,\ldots,\tau_m$ adaptively.
This would be achieved by adjusting  the method of \cite{exponentialBFNPW14}, 
but with the strong proviso of exact sparsity.

Secondly, 
it is worth asking to what extent the trade-offs between the different algorithms reflect reality.
In particular, is it only an artifact of the proof that the simpler algorithm based on hard thresholding applies to a narrower class of signals?

\subsection{Organization}
The remainder of the paper is organized as follows.  
In Section~\ref{SecTec}, we outline some technical tools upon which our results rely,
namely some properties of Gaussian random matrices . 
In Section~\ref{SecDir}, we consider recovery of the direction $\vf/\|\vf\|$  only and we propose two algorithms to achieve it.  
In Section~\ref{SecDirMag}, we present three algorithms for the recovery of the entire signal $\vf$.  
Finally, in Section~\ref{SecPfs}, we provide proofs for the results outlined in Section~\ref{SecTec}.

\section{Technical ingredients}
\label{SecTec}

In this section, we highlight the theoretical properties upon which our results rely.
Their proofs are deferred to Section \ref{SecPfs}
so that the reader does not lose track of our objectives.
The first property we put forward is an adaptation to the dictionary case of the so-called sign product embedding property
(the term was coined in \cite{jacques2013quantized} but the result originally appeared in \cite{pv-noisy-1bit}).

\bthm[$\vD$-SPEP]
\label{ThmSPEP}
Let $\delta > 0$,
let $m \ge C \delta^{-7} s \ln(eN/s)$,
and let $\vA \in \bR^{m \times n}$ be populated by independent standard normal random variables.
Then, with failure probability at most $\gamma \exp(-c \delta^2 m)$,
the renormalized matrix $\vA':= (\sqrt{2/\pi}/m) \vA$
satisfies the $s$th order sign product embedding property adapted to $\vD \in \bR^{n \times N}$ with constant $\delta$ --- $\vD$-SPEP$(s,\delta)$ for short ---
i.e., 
\be
\label{SPEP}
\left| 
 \langle \vA' \vf, \sgn(\vA' \vg) \rangle
- \langle \vf, \vg \rangle
\right| \le \delta
\ee
holds for all $\vf, \vg \in \vD(\Sigma^N_s) \cap S^{n-1}$.
\ethm

\brk
The power $\delta^{-7}$ is unlikely to be optimal.
At least in the nondictionary case, i.e., when $\vD = \vI_n$,
it can be reduced to $\delta^{-2}$, see \cite{bilyk2015random}.
\erk

As an immediate consequence of $\vD$-SPEP, 
setting $\vg = \vf$ in \eqref{SPEP} allows one to deduce a variation of the classical restricted isometry property adapted to $\vD$, where the inner norm becomes the $\ell_1$-norm
(we mention in passing that this variation could also be deduced by other means).

\bcor[$\vD$-RIP$_1$]
Let $\delta > 0$,
let $m \ge C \delta^{-7} s \ln(eN/s)$,
and let $\vA \in \bR^{m \times n}$ be populated by independent standard normal random variables.
Then, with failure probability at most $\gamma \exp(-c \delta^2 m)$,
the renormalized matrix $\vA':= (\sqrt{2/\pi}/m) \vA$
satisfies the $s$th-order  $\ell_1$-restricted isometry property adapted to $\vD \in \bR^{n \times N}$ with constant $\delta$
--- $\vD$-RIP$_{1}(s,\delta)$ for short ---
i.e.,
\be
(1-\delta) \| \vf\|_2 \le \| \vA' \vf \|_1 \le (1+\delta) \|\vf\|_2
\ee
holds for all $\vf \in \vD(\Sigma_s^N)$.
\ecor

The next property we put forward is an adaptation of the tessellation of the ``effectively sparse sphere'' (see \cite{pv-embeddings}) to the dictionary case.  
In what follows, given a (noninvertible) matrix $\vM$ and a set $K$, we denote by $\vM^{-1} (K)$ the preimage of $K$ with respect to $\vM$.

\bthm[Tessellation]
\label{ThmTesSphere}
Let $\eps > 0$,
let $m \ge C \eps^{-6} s \ln(eN/s)$,
and let $\vA \in \bR^{m \times n}$ be populated by independent standard normal random variables.
Then, with failure probability at most $\gamma \exp(-c \eps^2 m)$,
the rows $\va_1,\ldots,\va_m \in \bR^n$ of $\vA$ $\eps$-tessellate the effectively $s$-analysis-sparse sphere
--- we write that $\vA$ satisfies $\vD$-TES$(s,\eps)$ for short ---
i.e.,
\be
\label{Tes}
[ \vf,\vg \in (\vD^*)^{-1}(\Sigma_{s}^{N,{\rm eff}}) \cap S^{n-1} : \; \sgn \langle \va_i, \vf \rangle = \sgn \langle \va_i, \vg \rangle \mbox{ for all } i =1,\ldots,m  ]
\imp 
[ \|\vf - \vg\|_2 \le \eps ].
\ee
\ethm

\section{Signal estimation: direction only}
\label{SecDir}

In this whole section, given a measurement matrix $\vA \in \bR^{m \times n}$ with rows $\va_1,\ldots,\va_m \in \bR^n$,
the signals $\vf \in \bR^n$ are acquired via $\vy = \sgn(\vA \vf) \in \{-1,+1\}^m$,
i.e.,
$$
y_i = \sgn \langle \va_i, \vf \rangle,
\qquad 
i = 1,\ldots,m.
$$
Under this model, all $c \vf$ with $c>0$ produce the same one-bit measurements,
so one can only hope to recover the direction of $\vf$.
We present two methods to do so,
one based on linear programming and the other one based on hard thresholding.

\subsection{Linear programming}

Given a signal $\vf \in \bR^n$ observed via $\vy = \sgn (\vA \vf)$,
the optimization scheme we consider here consists in outputting the signal $\vf_{\rm lp}$ solution of
\be
\label{LPforDir}
\minimize{\vh \in \bR^n} \| \vD^* \vh\|_1
\qquad \mbox{subject to} \quad
\sgn( \vA \vh) = \vy,
\quad \|\vA \vh\|_1 = 1.
\ee
This is in fact a linear program (and thus may be solved efficiently), since the condition $\sgn(\vA \vh) = \vy$ reads
$$ 
y_i (\vA \vh)_i \ge 0
\qquad \mbox{for all } i = 1,\ldots, m,
$$
and, under this constraint, the condition $\|\vA \vh\|_1 = 1$ reads
$$
\sum_{i=1}^m y_i (\vA \vh)_i = 1.
$$

\bthm
\label{ThmLPfromSPEP}
If $\vA \in \bR^{m \times n}$ satisfies both $\vD$-TES$(36s,\eps)$ and $\vD$-RIP$_1(25s,1/5)$, 
then any effectively $s$-analysis-sparse signal $\vf \in (\vD^*)^{-1}\Sigma_s^{N,{\rm eff}}$
observed via $\vy = \sgn(\vA \vf)$ is directionally  approximated by the output $\vf_{\rm lp}$ of the linear program \eqref{LPforDir} with error
$$
\left\| \f{\vf}{\|\vf\|_2} - \f{\vf_{\rm lp}}{\|\vf_{\rm lp}\|_2} \right\|_2 \le \eps.
$$ 
\ethm

\bpf
The main step is to show that
$\vf_{\rm lp}$ is effectively $36s$-analysis-sparse when $\vD$-RIP$_1(t,\delta)$ holds with $t= 25s$ and $\delta=1/5$.
Then, since both $\vf/\|\vf\|_2$ and $\vf_{\rm lp} / \|\vf_{\rm lp}\|_2$ belong to $ (\vD^*)^{-1}\Sigma_{36 s}^{N,{\rm eff}} \cap S^{n-1}$ and have the same sign observations, 
$\vD$-TES$(36s,\eps)$ implies the desired conclusion.
To prove the effective analysis-sparsity of $\vf_{\rm lp}$, we first estimate $\|\vA \vf\|_1$ from below.
For this purpose, let $T_0$ denote an index set of $t$ largest absolute  entries of $\vD^* \vf$,
$T_1$ an index set of next $t$ largest absolute entries of $\vD^* \vf$, $T_2$ an index set of next $t$ largest absolute entries of $\vD^* \vf$,  etc..
 We have
 \begin{align*}
 \|\vA \vf \|_1 & = \|\vA \vD \vD^* \vf\|_1 
 = \left\| \vA \vD \left( \sum_{k \ge 0} (\vD^*\vf )_{T_k} \right) \right\|_1
  \ge \|\vA \vD \left( (\vD^* \vf)_{T_0} \right)\|_1
- \sum_{k \ge 1} \|\vA \vD \left( (\vD^* \vf)_{T_k} \right)\|_1\\
& \ge (1-\delta) \|\vD \left( (\vD^* \vf )_{T_0} \right)\|_2
- \sum_{k \ge 1} (1+\delta) \|\vD \left( (\vD^* \vf )_{T_k} \right)\|_2,
 \end{align*}
 where the last step used $\vD$-RIP$_1(t,\delta)$.
 We notice that, for $k \ge 1$,
 $$
 \|\vD \left( (\vD^* \vf )_{T_k} \right)\|_2 
 \le \| (\vD^* \vf )_{T_k} \|_2
 \le \f{1}{\sqrt{t}} \| (\vD^* \vf )_{T_{k-1}} \|_1,
 $$
 from where it follows that 
 \be
\label{LowerAf}
 \|\vA \vf\|_1 
 \ge (1-\delta) \|\vD \left( (\vD^* \vf )_{T_0}\right)\|_2 - \f{1+\delta}{\sqrt{t}} \|\vD^* \vf \|_1.
 \ee
 In addition, we observe that 
 \begin{align*}
 \|\vD^* \vf \|_2 & = \|\vf\|_2 = \|\vD \vD^* \vf\|_2
 = \left\| \vD \left( \sum_{k \ge 0} (\vD^* \vf)_{T_k} \right) \right\|_2
 \le \left\| \vD \left( (\vD^* \vf)_{T_0} \right) \right\|_2
 + \sum_{k \ge 1} \left\| \vD \left( (\vD^* \vf)_{T_k} \right) \right\|_2\\
 & \le \left\| \vD \left( (\vD^* \vf)_{T_0} \right) \right\|_2 + \f{1}{\sqrt{t}} \|\vD^* \vf \|_1.
 \end{align*}
 In view of the effective sparsity of $\vD^* \vf$, we obtain
 $$
 \|\vD^* \vf\|_1
 \le \sqrt{s} \|\vD^* \vf\|_2 \le 
 \sqrt{s}\left\| \vD \left( (\vD^* \vf)_{T_0} \right) \right\|_2 + \sqrt{s/t} \|\vD^* \vf \|_1,
 $$
 hence
 \be
\label{LowerDD*T0}
 \left\| \vD \left( (\vD^* \vf)_{T_0} \right) \right\|_2 \ge \f{1- \sqrt{s/t}}{\sqrt{s}} \|\vD^* \vf \|_1.
 \ee
 Substituting \eqref{LowerDD*T0} in \eqref{LowerAf} yields
 \be
\label{LowerAf2}
  \|\vA \vf\|_1  \ge \left( (1-\delta)(1-\sqrt{s/t}) - (1+\delta)(\sqrt{s/t}) \right) \f{1}{\sqrt{s}} \|\vD^* \vf\|_1
  = \f{2/5}{\sqrt{s}} \|\vD^* \vf\|_1,
 \ee
 where we have used the values $t = 25s$ and $\delta=1/5$.
This lower estimate for $\|\vA \vf \|_1$,
combined with the minimality property of $\vf_{\rm lp}$,
allows us to derive that 
\be
\label{UpperD*fhat}
\|\vD^* \vf_{\rm lp} \|_1
\le \|\vD^*( \vf/ \|\vA \vf\|_1 )\|_1 = \f{\|\vD^* \vf\|_1}{\|\vA \vf \|_1}
\le (5/2) \sqrt{s}.
\ee
Next, with $\wh{T}_0$ denoting an index set of $t$ largest absolute  entries of $\vD^* \vf_{\rm lp}$,
$\wh{T}_1$ an index set of next $t$ largest absolute entries of $\vD^* \vf_{\rm lp}$, $\wh{T}_2$ an index set of next $t$ largest absolute entries of $\vD^* \vf_{\rm lp}$,  etc., we can write
\begin{align*}
1 & = \|\vA \vf_{\rm lp} \|_1 = 
\|\vA \vD \vD^* \vf_{\rm lp} \|_1 = 
\left\| \vA \vD \left( \sum_{k \ge 0} (\vD^* \vf_{\rm lp})_{\wh{T}_k} \right) \right\|_1
\le \sum_{k \ge 0} \left\| \vA \vD \left(  (\vD^* \vf_{\rm lp})_{\wh{T}_k} \right) \right\|_1\\
& \le \sum_{k \ge 0} (1+\delta) \left\|  \vD \left( (\vD^* \vf_{\rm lp})_{\wh{T}_k} \right) \right\|_2 
 = (1+\delta) \left[ \left\|  \ (\vD^* \vf_{\rm lp})_{\wh{T}_0} \right\|_2
+  \sum_{k \ge 1} \left\|  \ (\vD^* \vf_{\rm lp})_{\wh{T}_k} \right\|_2 \right]\\
& \le (1+\delta) \left[ \|\vD^* \vf_{\rm lp} \|_2 + \f{1}{\sqrt{t}} \|\vD^* \vf_{\rm lp}\|_1 \right]
\le (1+\delta) \left[ \|\vD^* \vf_{\rm lp} \|_2 + (5/2)\sqrt{s/t} \right].
\end{align*}
This chain of inequalities shows that 
\be
\label{LowerD*fhat}
\|\vD^* \vf_{\rm lp} \|_2  \ge \f{1-(5/2)\sqrt{s/t}}{1+\delta} = \f{5}{12}.
\ee
Combining \eqref{UpperD*fhat} and \eqref{LowerD*fhat}, we obtain 
$$
\|\vD^* \vf_{\rm lp} \|_1 \le 6 \sqrt{s} \|\vD^* \vf_{\rm lp} \|_2.
$$
In other words, $\vD^* \vf_{\rm lp}$ is effectively $36s$-sparse, which is what was needed to conclude the proof.
\epf

\brk
We point out that if $\vf$ was genuinely, instead of effectively, $s$-analysis-sparse,
then a lower bound of the type \eqref{LowerAf2} would be immediate from the $\vD$-RIP$_1$.
We also point out that our method of proving that the linear program outputs an effectively analysis-sparse signal is new even in the  case $\vD = \vI_n$.
In fact, it makes it possible to remove a logarithmic factor from the number of measurements in this ``nondictionary'' case, too (compare with \cite{pv-1-bit}).  
Furthermore, it allows for an analysis of the linear program \eqref{LPforDir} only based on deterministic conditions that the matrix $\vA$ may satisfy.
\erk

\subsection{Hard thresholding}

Given a signal $\vf \in \bR^n$ observed via $\vy = \sgn (\vA \vf)$,
the hard thresholding scheme we consider here consists in constructing a signal $\vf_{\rm ht} \in \bR^n$ as
\be
\label{HTforDir}
\vf_{\rm ht} = \vD \vz,
\qquad \mbox{where }
\vz := H_t(\vD^* \vA^* \vy ).
\ee
Our recovery result holds for $s$-synthesis sparse signals that are also effectively $\kappa s$-analysis-sparse for some $\kappa \ge 1$
(we discussed in the introduction some choices of dictionaries $\vD$ making this happen).

\bthm
\label{ThmHTDir}
If $\vA \in \bR^{m \times n}$ satisfies $\vD$-SPEP$(s+t,\eps/8)$,
$t = \lceil 16 \eps^{-2} \kappa s \rceil$,
then any $s$-synthesis-sparse signal $\vf \in \vD(\Sigma_s^N)$ with $\vD^* \vf \in \Sigma_{\kappa s}^{N,{\rm eff}}$ observed via $\vy = \sgn (\vA \vf)$
is directionally approximated
by the output $\vf_{\rm ht}$ of the hard thresholding \eqref{HTforDir}
with error
$$
\left\| \f{\vf}{\|\vf\|_2} - \f{\vf_{\rm ht}}{\|\vf_{\rm ht}\|_2} \right\|_2 \le \eps.
$$
\ethm

\bpf 
We assume without loss of generality that $\|\vf\|_2 = 1$.
Let $T=T_0$ denote an index set of $t$ largest absolute  entries of $\vD^* \vf$,
$T_1$ an index set of next $t$ largest absolute entries of $\vD^* \vf$, $T_2$ an index set of next $t$ largest absolute entries of $\vD^* \vf$,  etc..
We start by noticing that $\vz$ is a better $t$-sparse approximation to
$\vD^* \vA^* \vy = \vD^* \vA^* \sgn(\vA \vf)$ than $[\vD^* \vf]_T$,
so we can write
$$
\| \vD^* \vA^* \sgn(\vA \vf) - \vz \|_2^2
\le
\|\vD^* \vA^* \sgn(\vA \vf) - [\vD^* \vf]_T \|_2^2,
$$
i.e.,
$$
\| (\vD^* \vf - \vz) - (\vD^* \vf - \vD^* \vA^* \sgn(\vA \vf)) \|_2^2
\le
\| (\vD^* \vf - \vD^* \vA^* \sgn(\vA \vf)) - [\vD^* \vf]_{\ol{T}} \|_2^2.
$$
Expanding the squares and rearranging gives
\begin{align}
\label{Term1}
\|\vD^* \vf - \vz \|_2^2 
& \le 2 \langle \vD^* \vf - \vz, \vD^* \vf - \vD^* \vA^* \sgn(\vA \vf) \rangle \\
\label{Term2}
& - 2 \langle [\vD^* \vf]_{\ol{T}} , \vD^* \vf - \vD^* \vA^* \sgn(\vA \vf) \rangle \\
\label{Term3}
& + \| [\vD^* \vf]_{\ol{T}} \|_2^2. 
\end{align}
To bound \eqref{Term3}, we invoke \cite[Theorem 2.5]{foucart2013} and the effective analysis-sparsity of $\vf$  to derive
$$
 \| [\vD^* \vf]_{\ol{T}} \|_2^2 
 \le \f{1}{4t}  \| \vD^* \vf \|_1^2
 \le \f{\kappa s}{4t} \| \vD^* \vf \|_2^2 
 = \f{\kappa s}{4t} \|\vf \|_2^2 
= \f{\kappa s}{4t}.
$$ 
To bound \eqref{Term1} in absolute value, we notice that it can be written
as
\begin{align*}
2 | \langle \vD \vD^* \vf - \vD \vz, &\vf - \vA^* \sgn(\vA \vf) \rangle |
=  2 | \langle \vf - \vf_{\rm ht}, \vf - \vA^* \sgn(\vA \vf) \rangle | \\
& = 2 | \langle \vf - \vf_{\rm ht}, \vf \rangle - \langle \vA (\vf - \vf_{\rm ht}),  \sgn(\vA \vf) \rangle |
\le 2 \eps' \|\vf - \vf_{\rm ht} \|_2,
\end{align*}
where the last step followed from $\vD$-SPEP$(s+t,\eps')$, $\eps' := \eps /8$.
Finally, \eqref{Term2} can be bounded in absolute value by
\begin{align*}
2 & \sum_{k \ge 1} | \langle [\vD^* \vf]_{T_k}, \vD^*(\vf - \vA^* \sgn(\vA \vf)) \rangle | = 
 2 \sum_{k \ge 1} | \langle \vD([\vD^* \vf]_{T_k}), \vf - \vA^* \sgn(\vA \vf) \rangle | \\
 & = 2 \sum_{k \ge 1} | \langle \vD([\vD^* \vf]_{T_k}), \vf \rangle -
 \langle \vA (\vD([\vD^* \vf]_{T_k})), \sgn(\vA \vf) \rangle |
  \le
  2 \sum_{k \ge 1} \eps' \| \vD([\vD^* \vf]_{T_k}) \|_2\\
&  \le 2 \eps' \sum_{k \ge 1} \| [\vD^* \vf]_{T_k} \|_2
  \le 2 \eps' \sum_{k \ge 1} \f{\| [\vD^* \vf]_{T_{k-1}} \|_1}{\sqrt{t}}
\le 2 \eps' \f{\|\vD^* \vf\|_1}{\sqrt{t}}
\le 2 \eps' \f{\sqrt{\kappa  s} \|\vD^* \vf\|_2}{\sqrt{t}}
= 2 \eps'  \sqrt{\f{\kappa s}{t}}.
\end{align*}
Putting everything together, we obtain
$$
\|\vD^* \vf - \vz \|_2^2  \le 2 \eps' \|\vf - \vf_{\rm ht}\|_2 + 2 \eps'  \sqrt{\f{\kappa s}{t}} + \f{\kappa s}{4t}.
$$
In view of 
$\|\vf - \vf_{\rm ht}\|_2 = \|\vD (\vD^* \vf - \vz) \|_2 \le \|\vD^* \vf - \vz\|_2$,
it follows that
$$
\|\vf - \vf_{\rm ht}\|_2^2  \le 2 \eps' \|\vf - \vf_{\rm ht}\|_2 + 2 \eps'  \sqrt{\f{\kappa s}{t}} + \f{\kappa s}{4t},
\quad \mbox{i.e., } \;
( \|\vf - \vf_{\rm ht}\|_2 - \eps' )^2 \le {\eps'}^2 + 2 \eps'  \sqrt{\f{ \kappa s}{t}} + \f{\kappa s}{4t}
\le \left( \eps' \hspace{-0.5mm}+\hspace{-0.5mm} \sqrt{\f{\kappa s}{t}} \right)^2 \hspace{-1mm}.
$$
This implies that
$$
 \|\vf - \vf_{\rm ht}\|_2 \le 2 \eps' +  \sqrt{\f{ \kappa s}{t}}.
$$
Finally, since $\vf_{\rm ht}/\|\vf_{\rm ht}\|_2$ is the best $\ell_2$-normalized approximation to $\vf_{\rm ht}$, we conclude that
$$
\left\| \vf - \f{\vf_{\rm ht}}{\|\vf_{\rm ht}\|_2} \right\|_2 
 \le 
 \|\vf - \vf_{\rm ht}\|_2  + \left\| \vf_{\rm ht} - \f{\vf_{\rm ht}}{\|\vf_{\rm ht}\|_2} \right\|_2
 \le 2  \|\vf - \vf_{\rm ht}\|_2
 \le 4 \eps' + 2 \sqrt{\f{ \kappa s}{t}}.
$$
The announced result follows from our choices of $t$ and $\eps'$.
\epf

\section{Signal estimation: direction and magnitude}
\label{SecDirMag}

Since information of the type $y_i = \sgn \langle \va_i,\vf \rangle$
can at best allow one to estimate the direction of a signal $\vf \in \bR^n$,
we consider in this section information of the type
$$
y_i = \sgn( \langle \va_i, \vf \rangle - \tau_i ),
\qquad i  = 1,\ldots,m ,
$$
for some thresholds $\tau_1,\ldots,\tau_m$ introduced before quantization.
In the rest of this section, we give three methods for recovering $\vf$ in its entirety.  
The first one is based on linear programming, 
the second one on second-order code programming,
and the last one on hard thresholding.

We are going to show that using these algorithms, one can  estimate both the direction and the magnitude of dictionary-sparse signal $\vf \in \bR^n$
given a prior magnitude bound  such as  $\| \vf \|_2 \le r$.
We simply rely on the previous results by ``lifting'' the situation from $\bR^n$ to $\bR^{n+1}$,
in view of the observation that
$\vy = \sgn (\vA \vf - \boldsymbol{\tau})$
can be interpreted as
$$
\vy = \sgn (\wt{\vA} \wt{\vf}),
\qquad \mbox{where} \quad
\wt{\vA} :=  \bbmx 
& & \vline & -\tau_1/c\\
& \vA & \vline & \vdots\\
& & \vline & -\tau_m/c
\ebmx \in \bR^{m \times (n+1)},
\quad
\wt{\vf} := \bbmx  \vf \\ \hline c \ebmx \in \bR^{n+1}.
$$
The following lemma will be equally useful when dealing with linear programming,
second-order cone programming, or with
 hard thresholding schemes.

\blem
\label{LemTrueAugm}
For $\wt{\vf}, \wt{\vg} \in \bR^{n+1}$ written as
$$
\wt{\vf} := \bbmx  \vf_{[n]} \\ \hline f_{n+1} \ebmx 
\qquad \mbox{ and } \qquad
\wt{\vg} =: \bbmx \vg_{[n]} \\ \hline g_{n+1} \ebmx
$$
with $\wt{\vf}_{[n]}, \wt{\vg}_{[n]} \in \bR^n$ and with
 $f_{n+1} \not= 0$, $g_{n+1} \not= 0$,
 one has
$$
\left\| \f{\vf_{[n]}}{f_{n+1}} - \f{\vg_{[n]}}{g_{n+1}} \right\|_2
\le \f{\|\wt{\vf}\|_2 \|\wt{\vg}\|_2}{|f_{n+1}||g_{n+1}|}
\left\| \f{\wt{\vf}}{\|\wt{\vf}\|_2} - \f{\wt{\vg}}{\|\wt{\vg}\|_2} \right\|_2.
$$
\elem

\bpf
By using the triangle inequality in $\bR^n$ and Cauchy--Schwarz inequality in $\bR^2$, we can write
\begin{align*}
\left\| \f{\vf_{[n]}}{f_{n+1}} - \f{\vg_{[n]}}{g_{n+1}} \right\|_2
& = \|\wt{\vf}\|_2 \left\| \f{1/f_{n+1}}{\|\wt{\vf}\|_2} \vf_{[n]} - \f{1/g_{n+1}}{\|\wt{\vf}\|_2} \vg_{[n]} \right\|_2\\
 & \le \|\wt{\vf}\|_2 \left( \f{1}{f_{n+1}}
\left\| \f{\vf_{[n]}}{\| \wt{\vf} \|_2} - \f{\vg_{[n]}}{\|\wt{\vg}\|_2} \right\|_2
+ \left| \f{1/g_{n+1}}{\|\wt{\vf}\|_2} - \f{1/f_{n+1}}{\|\wt{\vg}\|_2} \right| \|\vg_{[n]}\|_2
\right)\\
& = 
\|\wt{\vf}\|_2 \left( \f{1}{f_{n+1}}
\left\| \f{\vf_{[n]}}{\| \wt{\vf} \|_2} - \f{\vg_{[n]}}{\|\wt{\vg}\|_2} \right\|_2
+ \f{\|\vg_{[n]}\|_2}{|f_{n+1}| |g_{n+1}|}
\left| \f{f_{n+1}}{\|\wt{\vf}\|_2} - \f{g_{n+1}}{\|\wt{\vg}\|_2} \right| 
\right)\\
& \le \|\wt{\vf}\|_2
\left[ \f{1}{|f_{n+1}|^2} + \f{\|\vg_{[n]}\|_2^2}{|f_{n+1}|^2 |g_{n+1}|^2} \right]^{1/2}
\left[ \left\| \f{\vf_{[n]}}{\| \wt{\vf} \|_2} - \f{\vg_{[n]}}{\|\wt{\vg}\|_2} \right\|_2^2
+ \left| \f{f_{n+1}}{\|\wt{\vf}\|_2} - \f{g_{n+1}}{\|\wt{\vg}\|_2} \right|^2
\right]^{1/2}\\
& =  \|\wt{\vf}\|_2
\left[ \f{\|\wt{\vg}\|_2^2}{|f_{n+1}|^2 |g_{n+1}|^2} \right]^{1/2}
\left\| \f{\wt{\vf}}{\|\wt{\vf}\|_2} - \f{\wt{\vg}}{\|\wt{\vg}\|_2} \right\|_2,
\end{align*}
which is the announced result.
\epf

\subsection{Linear programming}

Given a signal $\vf \in \bR^n$ observed via $\vy = \sgn (\vA \vf - \boldsymbol{\tau})$
with $\tau_1,\ldots,\tau_m \sim \cN(0,\sigma^2)$,
the optimization scheme we consider here consists in outputting the signal
\be
\label{Defflp}
\vf_{\rm LP} = \f{\sigma}{\wh{u}} \wh{\vh} \in \bR^n,
\ee
where $\wh{\vh} \in \bR^{n}$ and $\wh{u} \in \bR$
are solutions of
\be
\label{OptProg}
\underset{\vh \in \bR^n, u \in \bR}{\rm minimize \;} \;
\|\vD^* \vh \|_1 + |u|
\qquad \mbox{subject to} \quad
\sgn(\vA \vh - u \boldsymbol{\tau} / \sigma) = \vy,  
\quad \|\vA \vh - u \boldsymbol{\tau} / \sigma \|_1 = 1.
\ee

\bthm
\label{ThmDirMagLP}
Let $\eps, r, \sigma > 0$, 
let $m \ge C (r/\sigma+\sigma/r)^6 \eps^{-6} s \ln(eN/s)$,
and let $\vA \in \bR^{m \times n}$ be populated by independent standard normal random variables.
Furthermore, let $\tau_1,\ldots,\tau_m$ be independent normal random variables with mean zero and variance $\sigma^2$
that are also independent from the entries of $\vA$.
Then, with failure probability at most $\gamma \exp(-c m \eps^2 r^2 \sigma^2/(r^2+\sigma^2)^2)$,
any effectively $s$-analysis sparse $\vf \in \bR^n$
satisfying $\|\vf\|_2 \le r$
and observed via $\vy = \sgn(\vA \vf - \boldsymbol{\tau})$
is approximated by
$\vf_{\rm LP}$ given in \eqref{Defflp}
 with error
$$
\left\| \vf- \vf_{\rm LP} \right\|_2 \le  \eps r.
$$
\ethm

\bpf
Let us introduce the ``lifted'' signal $\wt{\vf} \in \bR^{n+1}$,
the ``lifted'' tight frame $\wt{\vD} \in \bR^{(n+1)\times (N+1)}$,
and the ``lifted'' measurement matrix $\wt{\vA} \in \bR^{m \times (N+1)}$ defined as
\be
\label{LiftedObj}
\wt{\vf} := \bbmx  \vf \\ \hline \sigma \ebmx,
\qquad
\wt{\vD} := \bbmx
\, \vD & \vline & {\bf 0}\, \\ \hline \, {\bf 0} & \vline & 1 \,
\ebmx,
\qquad
\wt{\vA} :=  \bbmx 
& & \vline & -\tau_1/\sigma\\
& \vA & \vline & \vdots\\
& & \vline & -\tau_m/\sigma
\ebmx .
\ee
First, we observe that $\wt{\vf}$ is effectively $(s+1)$-analysis-sparse (relative to $\wt{\vD}$),
since $\wt{\vD}^* \wt{\vf} = \bbmx \vD^* \vf \\ \hline \sigma \ebmx$, hence
$$
\f{\|\wt{\vD}^* \wt{\vf}\|_1}{\|\wt{\vD}^* \wt{\vf}\|_2}
= \f{\|\vD^* \vf \|_1 + \sigma}{\sqrt{\|\vD^* \vf\|_2^2+\sigma^2}}
\le \f{\sqrt{s} \|\vD^* \vf\|_2 + \sigma}{\sqrt{\|\vD^* \vf\|_2^2+\sigma^2}}
\le \sqrt{s+1}.
$$ 
Next, we observe that the matrix $\wt{\vA} \in \bR^{m \times (n+1)}$,
 populated by independent standard normal random variables,
satisfies $\wt{\vD}$-TES$(36(s+1),\eps')$,
$\eps' := \df{r \sigma}{2(r^2 + \sigma^2)} \eps$,
and $\wt{\vD}$-RIP$_1(25(s+1),1/5)$
with failure probability at most $\gamma \exp(-c m {\eps'}^2) + \gamma' \exp(-c' m)
\le \gamma'' \exp(-c'' m \eps^2 r^2 \sigma^2 / (r^2 + \sigma^2)^2)$,
since 
$m \ge C {\eps'}^{-6} (s+1) \ln(eN/(s+1))$
and
$m \ge C (1/5)^{-7} (s+1) \ln(e N / (s+1))$
are ensured by our assumption on $m$. 
Finally, we observe that $\vy = \sgn(\wt{\vA} \wt{\vf})$
and that the optimization program \eqref{OptProg} reads
$$
\underset{\wt{\vh} \in \bR^{n+1}}{\rm minimize \;}
\|\wt{\vD}^* \wt{\vh} \|_1
\qquad \mbox{subject to} \quad
\sgn(\wt{\vA} \wt{\vh} ) = \vy,  
\quad \|\wt{\vA} \wt{\vh}  \|_1 = 1.
$$
Denoting its solution as $\wt{\vg} = \bbmx \vg_{[n]} \\ \hline g_{n+1} \ebmx = \bbmx \wh{\vh} \\ \hline \wh{u} \ebmx \in \bR^{n+1}$,
Theorem \ref{ThmLPfromSPEP}
implies that 
$$
\left\| \f{\wt{\vf}}{\|\wt{\vf}\|_2} - \f{\wt{\vg}}{\|\wt{\vg}\|_2} \right\|_2 \le \eps'.
$$
In particular, looking at the last coordinate, this inequality yields
$$
\left| \f{\sigma}{\|\wt{\vf}\|_2} - \f{g_{n+1}}{\|\wt{\vg}\|_2} \right| \le \eps',
\qquad \mbox{hence} \qquad
\f{|g_{n+1}|}{\|\wt{\vg}\|_2}
\ge \f{\sigma}{\|\wt{\vf}\|_2} - \eps' 
\ge \f{\sigma}{\sqrt{r^2+\sigma^2}} - \f{\sigma}{2 \sqrt{r^2 + \sigma^2}}
= \f{\sigma}{2 \sqrt{r^2 + \sigma^2}}. 
$$
In turn, applying Lemma \ref{LemTrueAugm} while taking  $\vf = \vf_{[n]}$ and $\vf_{\rm LP} = (\sigma/g_{n+1}) \vg_{[n]}$ into consideration gives
$$
\left\| \f{\vf}{\sigma} - \f{\vf_{\rm LP}}{\sigma} \right\|_2
\le \f{\| \wt{\vf} \|_2}{\sigma} \f{\| \wt{\vg} \|_2}{|g_{n+1}|} \eps'
\le \f{\| \wt{\vf} \|_2}{\sigma} \f{2\sqrt{r^2+\sigma^2}}{\sigma} \f{r \sigma}{2(r^2 + \sigma^2)} \eps
= \f{\| \wt{\vf} \|_2}{\sigma} \f{r}{\sqrt{r^2+\sigma^2}}
 \eps,
$$
so that 
$$
\| \vf - \vf_{\rm LP} \|_2 \le \|\wt{\vf}\|_2 \f{r}{\sqrt{r^2+\sigma^2}}  \eps
 \le r \eps .
$$
This establishes the announced result.
\epf

\brk
The recovery scheme \eqref{OptProg} does not require an estimation of $r$ to be run.
The recovery scheme presented next does require such an estimation.
Moreover, it is a second-order cone program instead of a simpler linear program.
But it has one noticeable advantage, 
namely that it not only applies to signals satisfying $\|\vD^* \vf\|_1 \le \sqrt{s}\|\vD^*\vf\|_2$
and $\|\vD^*\vf\|_2 \le r$
but more generally to 
signals satisfying $\|\vD^* \vf\|_1 \le \sqrt{s} r$ and $\|\vD^*\vf\|_2 \le r$.
For both schemes, one needs $\sigma$ to be of the same order as~$r$ for the results to become meaningful in terms of number of measurement and success probability.
However, if $r$ is only upper-estimated,
then one could choose $\sigma \ge r$ and obtain a weaker recovery error $\|\vf - \wh{\vf}\|_2 \le \eps \sigma$
with relevant number of measurement and success probability.
\erk

\subsection{Second-order cone programming}

Given a signal $\vf \in \bR^n$  observed via $\vy = \sgn (\vA \vf - \boldsymbol{\tau})$
with $\tau_1,\ldots,\tau_m \sim \cN(0,\sigma^2)$,
the optimization scheme we consider here consists in outputting the signal
\be
\label{Deffcp}
\vf_{\rm CP} = \underset{\vh \in \bR^n}{\rm \argmin \;} \;
\|\vD^* \vh\|_1
\qquad \mbox{subject to} \quad
\sgn(\vA \vh - \boldsymbol{\tau}) = \vy,
\quad \|\vh\|_2 \le r.
\ee

\bthm
\label{ThmSOCP}
Let $\eps, r, \sigma > 0$,
let $m \ge C (r/\sigma + \sigma/r)^6(r^2/\sigma^2+1) \eps^{-6} s \ln(eN/s)$,
and let $\vA \in \bR^{m \times n}$ be populated by independent standard normal random variables.
Furthermore, let $\tau_1,\ldots,\tau_m$ be independent normal random variables with mean zero and variance $\sigma^2$
that are also independent from $\vA$.
Then, with failure probability at most $\gamma \exp(- c' m \eps^2 r^2 \sigma^2 / (r^2+\sigma^2)^2)$,
any signal $\vf \in \bR^n$ with $\|\vf\|_2 \le r$, $\|\vD^* \vf\|_1 \le  \sqrt{s} r$,
and observed via $\vy = \sgn(\vA \vf - \boldsymbol{\tau})$
is approximated by 
$\vf_{\rm CP}$ given in \eqref{Deffcp}
 with error
$$
\left\| \vf- \vf_{\rm CP} \right\|_2 \le  \eps r.
$$
\ethm

\bpf
We again use the notation \eqref{LiftedObj} introducing  the ``lifted'' objects $\wt{\vf}$, $\wt{\vD}$, and $\wt{\vA}$.
Moreover, we set $\wt{\vg} := \bbmx  \vf_{\rm CP} \\ \hline \sigma  \ebmx$.
We claim that $\wt{\vf}$ and $\wt{\vg}$ are effectively $s'$-analysis-sparse,
$s' := ( r^2 / \sigma^2 + 1)( s+1)$.
For $\wt{\vg}$,
this indeed follows from
$ \| \wt{\vD}^* \wt{\vg} \|_2 = \|\wt{\vg}\|_2 = \sqrt{\|\vf_{\rm CP}\|_2^2 + \sigma^2} \ge \sigma$ and
$$
\|\wt{\vD}^* \wt{\vg}\|_1 = \left\| \bbmx \vD^* \vf_{\rm CP} \\ \hline \sigma  \ebmx \right\|_1
= \| \vD^* \vf_{\rm CP}\|_1 + \sigma
\le \| \vD^* \vf \|_1 + \sigma
\le \sqrt{s} r + \sigma \le \sqrt{r^2 + \sigma^2} \sqrt{s+1}.  
$$
We also notice that $\wt{\vA}$
satisfies $\wt{\vD}$-TES$(s', \eps')$,
$\eps' := \df{r \sigma}{r^2 + \sigma^2} \eps$,
with failure probability at most $\gamma \exp(-c m {\eps'}^2) \le \gamma \exp(- c' m \eps^2 r^2 \sigma^2 / (r^2+\sigma^2)^2)$,
since $m \ge C {\eps'}^{-6} s' \ln(eN/s')$
is ensured by our assumption on $m$.
Finally, we observe that both $\wt{\vf}/ \|\wt{\vf}\|_2$ and $\wt{\vg}/ \|\wt{\vg}\|_2$
are  $\ell_2$-normalized effectively $s'$-analysis-sparse
 and have the same sign observations $\sgn(\wt{\vA} \wt{\vf}) = \sgn(\wt{\vA} \wt{\vg}) = \vy$.
Thus,
$$
\left\| \f{\wt{\vf}}{\|\wt{\vf}\|_2} - \f{\wt{\vg}}{\|\wt{\vg}\|_2} \right\|_2 \le \eps'.
$$
In view of Lemma \ref{LemTrueAugm},
we derive
$$
\left\| \f{\vf}{\sigma} - \f{\vf_{\rm CP}}{\sigma} \right\|_2 
\le \f{r^2 + \sigma^2}{\sigma^2} \eps',
\qquad \mbox{hence} \qquad
\|\vf - \vf_{\rm CP}\|_2 \le \f{r^2 + \sigma^2}{\sigma} \eps' = r \eps. 
$$
This establishes the announced result.
\epf

\subsection{Hard thresholding}

Given a signal $\vf \in \bR^N$ observed via $\vy = \sgn(\vA \vf - \boldsymbol{\tau})$ with $\tau_1,\ldots,\tau_m \sim \cN(0,\sigma^2)$,
the hard thresholding scheme we consider here consists in  outputting the signal 
\be
\label{fht}
\vf_{\rm HT} = \f{-\sigma^2}{\langle \boldsymbol{\tau}, \vy \rangle} \vD \vz,
\qquad
\vz = H_{t-1}(\vD^* \vA^* \vy).
\ee

\bthm
Let $\eps, r, \sigma > 0$,
let $m \ge C \kappa  (r/\sigma+\sigma/r)^9 \eps^{-9} s \ln(eN/s)$,
and let $\vA \in \bR^{m \times n}$ be populated by independent standard normal random variables.
Furthermore, let $\tau_1,\ldots,\tau_m$ be independent normal random variables with mean zero and variance $\sigma^2$
that are also independent from the entries of $\vA$.
Then, with failure probability at most $\gamma \exp(-c m \eps^2 r^2 \sigma^2/(r^2+\sigma^2)^2)$,
any $s$-synthesis sparse and effectively $\kappa s$-analysis sparse signal $\vf \in \bR^n$
satisfying $\|\vf\|_2 \le r$
and observed via $\vy = \sgn(\vA \vf - \boldsymbol{\tau})$
is approximated by
$\vf_{\rm HT}$ given in \eqref{fht}  for $t:=\lceil 16 (\eps'/8)^{-2} \kappa (s+1) \rceil$ with error 
$$
\left\| \vf- \vf_{\rm HT} \right\|_2 \le  \eps r.
$$
\ethm

\bpf
We again use the notation  \eqref{LiftedObj} for the ``lifted'' objects $\wt{\vf}$, $\wt{\vD}$,
and $\wt{\vA}$.
First, we notice that $\wt{\vf}$ is $(s+1)$-synthesis sparse (relative to $\wt{\vD}$),
as well as effectively $\kappa (s+1)$-analysis sparse,
since $\wt{\vD}^* \wt{\vf} = \bbmx \vD^* \vf \\ \hline \sigma \ebmx$ satisfies
$$
\f{\|\wt{\vD}^* \wt{\vf}\|_1}{\|\wt{\vD}^* \wt{\vf}\|_2}
= \f{\|\vD^* \vf\|_1 + \sigma}{\sqrt{ \|\vD^*\vf\|_2^2 + \sigma^2 }}
\le  \f{\sqrt{\kappa s}\|\vD^* \vf\|_2 + \sigma}{\sqrt{ \|\vD^*\vf\|_2^2 +\sigma^2 }}
\le \sqrt{\kappa s + 1} \le \sqrt{\kappa (s+1)}.
$$
Next, we observe that the matrix $\wt{\vA} $,
populated by independent standard normal random variables,
satisfies $\wt{\vD}$-SPEP$(s+1+t,\eps'/8)$,
$\eps ' := \df{r \sigma}{2(r^2 + \sigma^2)} \eps$,
with failure probability at most $\gamma \exp(-c m {\eps'}^2 r^2)$,
since $m \ge C (\eps'/8)^{-7} (s+1+t) \ln(e(N+1)/(s+1+t))$
is ensured by our assumption on $m$.
Finally, since $\vy = \sgn(\wt{\vA} \wt{\vf})$,
Theorem \ref{ThmHTDir} implies that
$$
\left\| \f{\wt{\vf}}{\|\wt{\vf}\|_2} - \f{\wt{\vg}}{\|\wt{\vg}\|_2} \right\|_2 \le \eps',
$$
where $\wt{\vg} \in \bR^{n+1}$ is the output of the ``lifted'' hard thresholding scheme. i.e.,
$$
\wt{\vg} = \wt{\vD} \wt{\vz},
\qquad
\wt{\vz} = H_{t} (\wt{\vD}^*\wt{\vA}^* \vy),
$$
In particular, looking at the last coordinate, this inequality yields
\be
\label{LBg}
\left| \f{\sigma}{\|\wt{\vf}\|_2} - \f{g_{n+1}}{\|\wt{\vg}\|_2} \right| \le \eps',
\quad \mbox{hence} \quad
\f{|g_{n+1}|}{\|\wt{\vg}\|_2}
\ge \f{\sigma}{\|\wt{\vf}\|_2} - \eps' 
\ge \f{\sigma}{\sqrt{r^2+\sigma^2}} - \f{\sigma}{2 \sqrt{r^2 + \sigma^2}}
= \f{\sigma}{2 \sqrt{r^2 + \sigma^2}}. 
\ee
Now let us also observe that 
$$
\wt{\vz} = H_{t} \left( \bbmx \vD^* \vA^* \vy \\ \hline - \langle \boldsymbol{\tau},\vy \rangle / \sigma \ebmx \right)
 = \left\{ 
 \bmx
 \bbmx
H_{t}(\vD^* \vA^* \vy) \\ \hline 0
\ebmx, \\
\mbox{or \hspace{30mm}}\\
 \bbmx
H_{t-1}(\vD^* \vA^* \vy) \\ \hline - \langle \boldsymbol{\tau},\vy \rangle / \sigma
\ebmx ,
\emx
\right.
\quad \mbox{hence} \quad
\wt{\vg} = \wt{\vD} \wt{\vz} =
\left\{ 
 \bmx
 \bbmx
\vD(H_{t}(\vD^* \vA^* \vy)) \\ \hline 0
\ebmx, \\
\mbox{or \hspace{35mm}}\\
 \bbmx
\vD(H_{t-1}(\vD^* \vA^* \vy) ) \\ \hline - \langle \boldsymbol{\tau},\vy \rangle / \sigma
\ebmx .
\emx
\right.
$$
In view of \eqref{LBg}, the latter option prevails.
It is then apparent that $\vf_{\rm HT} = \sigma \vg_{[n]} / g_{n+1}$.
Lemma~\ref{LemTrueAugm} gives
$$
\left\| \f{\vf}{\sigma} - \f{\vf_{\rm HT}}{\sigma} \right\|_2
\le \f{\| \wt{\vf} \|_2}{\sigma} \f{\| \wt{\vg} \|_2}{|g_{n+1}|} \eps'
\le \f{\| \wt{\vf} \|_2}{\sigma} \f{2\sqrt{r^2+\sigma^2}}{\sigma} \f{r \sigma}{2(r^2 + \sigma^2)} \eps
= \f{\| \wt{\vf} \|_2}{\sigma} \f{r}{\sqrt{r^2+\sigma^2}}
 \eps,
$$
so that 
$$
\| \vf - \vf_{\rm HT} \|_2 \le \|\wt{\vf}\|_2 \f{r}{\sqrt{r^2+\sigma^2}}  \eps
 \le r \eps .
$$
This establishes the announced result.
\epf

\section{Postponed proofs and further remarks}\label{SecPfs}

This final section contains the theoretical justification of the technical properties underlying our results, followed by a few points of discussion around them.

\subsection{Proof of $\vD$-SPEP}

The Gaussian width turns out to be a useful tool in our proofs.
For a set $K \inc \bR^n$, it is defined by
$$
w(K) = \bE \left[ \sup_{\vf \in K } \langle \vf, \vg \rangle \right],
\qquad \vg \in \bR^n \mbox{ is a standard normal random vector}.
$$
We isolate the following two properties.

\blem
\label{LemGW}
Let $K \inc \bR^n$ be a linear space  and $K_1,\ldots,K_L \inc \bR^n$ be subsets of the unit sphere~$S^{n-1}$.
\begin{enumerate}\vspace{-5mm}
\item[(i)] $k / \sqrt{k+1} \le w(K \cap S^{n-1}) \le \sqrt{k}, \qquad k:= \dim (K)$;
\item[(ii)] $\displaystyle{w \left(  K_1 \cup \ldots \cup K_L \right) \le \max \left\{  w(K_1),\ldots,w(K_L) \right\} + 3 \sqrt{\ln(L)}}$.
\end{enumerate}
\elem

\bpf
(i) By the invariance under orthogonal transformation (see \cite[Proposition~2.1]{pv-noisy-1bit}\footnote{strictly speaking, \cite[Proposition~2.1]{pv-noisy-1bit} applies to the slightly different notion of mean width defined as $\bE \left[ \sup_{\vf \in K - K } \langle \vf, \vg \rangle \right]$}), we can assume that 
$K = \bR^k \times \left\{(0,\ldots,0) \right\}$.
We then notice that $\sup_{\vf \in K \cap S^{n-1}} \langle \vf, \vg \rangle = \|(g_1,\ldots,g_k)\|_2$
is the $\ell_2$-norm of a standard normal random vector of dimension $k$.
We invoke e.g. \cite[Proposition~8.1]{foucart2013} to derive the announced result.

(ii) Let us introduce the nonnegative random variables
$$
\xi_\ell := \sup_{\vf \in K_\ell} \langle \vf , \vg \rangle,
\quad \ell = 1,\ldots, L ,
$$
so that the Gaussian widths of each $K_\ell$ and of their union take the form
$$
w(K_\ell) = \bE(\xi_\ell),
\quad \ell = 1,\ldots, L,
\qquad \mbox{and} \qquad
w \left( K_1 \cup \cdots \cup K_L \right) = \bE \left( \max_{\ell = 1, \ldots, L } \xi_\ell \right).
$$
By the concentration of measure inequality (see e.g. \cite[Theorem 8.40]{foucart2013}) 
applied to the function $F: \vx \in \bR^n \mapsto \sup_{\vf \in K_\ell} \langle \vf, \vx \rangle$, which is a  Lipschitz function with  constant $1$,
each $\xi_\ell$ satisfies 
$$
\bP(\xi_\ell \ge \bE(\xi_\ell) + t ) 
\le \exp \left( -t^2/2 \right).
$$
Because each $\bE(\xi_\ell)$ is no larger than $\max_\ell \bE(\xi_\ell) = \max_\ell w(K_\ell) =: \omega$,
we also have
$$
\bP (\xi_\ell \ge \omega + t) \le \exp \left( -t^2/2 \right).
$$
Setting $v:= \sqrt{2 \ln(L)}$, we now calculate 
\begin{align*}
\bE \left( \max_{\ell =1,\ldots, L } \xi_\ell  \right)
& = \int_0^\infty \bP \left( \max_{\ell =1,\ldots, L } \xi_\ell  \ge u \right) du
 = \left( \int_0^{\omega+v} + \int_{\omega+v}^\infty  \right) \bP \left( \max_{\ell = 1,\ldots, L } \xi_\ell \ge  u \right) du\\
& \le \int_0^{\omega + v} 1 du + \int_{\omega + v}^\infty \sum_{\ell=1}^L \bP \left(  \xi_\ell \ge  u \right) du
 = \omega + v + \sum_{\ell=1}^L \int_v^\infty \bP \left(  \xi_\ell \ge  \omega + t \right) dt\\
&
\le \omega + v  + L \int_v^\infty \exp \left( -t^2/2 \right) dt
 \le \omega + v + L \f{\exp(-v^2/2)}{v} \\
& = \omega + \sqrt{2 \ln(L)} + L \f{1/L}{\sqrt{2 \ln(L)}}
\le \omega + c \sqrt{\ln(L)},
\end{align*}
where $c=\sqrt{2} + (\sqrt{2} \ln(2))^{-1} \le 3$.
We have shown that 
$w \left(  K_1 \cup \ldots \cup K_L \right) \le \max_{\ell } w(K_\ell) + 3 \sqrt{\ln(L)}$,
as desired.
\epf

We now turn our attention to proving the awaited theorem.

\bpf[Proof of Theorem \ref{ThmSPEP}]
 According to \cite[Proposition 4.3]{pv-noisy-1bit}, 
 with $\vA' := (\sqrt{2/\pi}/m) \vA$,
 we have
$$
\left| \langle \vA' \vf, \sgn( \vA' \vg ) \rangle
- \langle \vf, \vg \rangle
\right| \le \delta,
$$
for all $\vf,\vg \in \vD(\Sigma_s^N) \cap S^{n-1}$
provided $m \ge C \delta^{-7} w(\vD(\Sigma_s^N) \cap S^{n-1})^2$,
so it is enough to upper-bound $w(\vD(\Sigma_s^N) \cap S^{n-1})$ appropriately.
To do so, with $\Sigma_S^N$ denoting the space $\{ \vx \in \bR^N: {\rm supp}(\vx) \inc S \}$ for any $S \inc \{ 1,\ldots, N \}$, we use Lemma \ref{LemGW} to write
\begin{align*}
w(\vD(\Sigma_s^N) \cap S^{n-1}) 
& = w \bigg(\bigcup_{|S|=s} \left\{ \vD(\Sigma_S^N) \cap S^{n-1} \right\} \bigg)
\underset{(ii)}{\le} \max_{|S|=s} w(\vD(\Sigma_S^N) \cap S^{n-1} ) + 3 \sqrt{ \ln \left( \binom{N}{s} \right)}\\
& \underset{(i)}{\le} \sqrt{s} + 3 \sqrt{ s \ln \left( eN/s \right)}
\le 4 \sqrt{s \ln \left( eN/s \right)}.
\end{align*}
The result is now immediate.
\epf

\subsection{Proof of TES}

We propose two approaches for proving  Theorem \ref{ThmTesSphere}.
One uses again the notion of Gaussian width,
the other one relies on covering numbers.
The necessary results are isolated in the following lemma.

\blem
\label{LemWNana}
The set of $\ell_2$-normalized effectively $s$-analysis-sparse signals satisfies
\begin{enumerate}[(i)]\vspace{-5mm}
\item $\displaystyle{w \left( (\vD^*)^{-1}(\Sigma_s^{N,{\rm eff}}) \cap S^{n-1} \right) } \le C \sqrt{s \ln(eN/s)},$
\item $\displaystyle{\cN \left( (\vD^*)^{-1}(\Sigma_s^{N,{\rm eff}}) \cap S^{n-1} , \rho \right)
\le \binom{N}{t}\left( 1 + \f{8}{\rho} \right)^t,
\qquad t := \lceil 4 \rho^{-2}} s \rceil .$
\end{enumerate}
\elem

\bpf
(i) By the definition of the Gaussian width for $\cK_s: = (\vD^*)^{-1}(\Sigma_s^{N,{\rm eff}}) \cap S^{n-1}$, with $\vg \in \bR^n$ denoting a standard normal random vector, 
\begin{equation}\label{slep}
w(\cK_s)  = \bE \left[ \sup_{\substack{\vD^* \vf \in \Sigma_s^{N,{\rm eff}} \\ \|\vf\|_2 = 1}}
\langle \vf , \vg \rangle \right]
=  \bE \left[ \sup_{\substack{\vD^* \vf \in \Sigma_s^{N,{\rm eff}} \\ \|\vD^* \vf\|_2 = 1}}
\langle \vD \vD^* \vf , \vg \rangle \right]
\le 
\bE \left[ \sup_{\substack{\vx \in \Sigma_s^{N,{\rm eff}} \\ \|\vx\|_2 = 1}}
\langle \vD \vx ,  \vg \rangle \right].
\end{equation}
In view of $\|\vD\|_{2 \to 2} = 1$, we have, for any $\vx,\vx' \in \Sigma_s^{N,{\rm eff}}$ with $\|\vx\|_2 = \|\vx'\|_2 =1$,
\begin{align*}
\bE \left( \langle \vD \vx, \vg \rangle - \langle \vD \vx', \vg' \rangle \right)^2
&= \bE \left[ \langle \vD \vx, \vg \rangle ^2 \right] + \bE \left[ \langle \vD \vx', \vg' \rangle ^2 \right]
= \|\vD \vx\|_2^2 + \|\vD \vx'\|_2^2  
\le \|\vx\|_2^2 + \|\vx'\|_2^2\\
&= \bE \left( \langle \vx, \vg \rangle - \langle \vx', \vg' \rangle \right)^2.
\end{align*}
Applying Slepian's lemma (see e.g. \cite[Lemma 8.25]{foucart2013}), we obtain
$$
w(\cK_s)  \le 
\bE \left[ \sup_{\substack{\vx \in \Sigma_s^{N,{\rm eff}} \\ \|\vx\|_2 = 1}}
\langle \vx ,  \vg \rangle \right]
 =w(\Sigma_s^{N,{\rm eff}} \cap S^{n-1}).
$$
The latter is known to be bounded by $C s \ln (eN/s)$, see \cite[Lemma~2.3]{pv-noisy-1bit}.

(ii) The covering number $\cN(\cK_s,\rho)$ is bounded above by the maximal number $\cP(\cK_s,\rho)$
of elements in $\cK_s$ that are separated by a distance $\rho$.
We claim that  
$\cP (\cK_s, \rho) \le \cP(\Sigma_t^N \cap B_2^N, \rho/2)$.
To justify this claim, let us consider a maximal $\rho$-separated set $\{\vf^1,\ldots,\vf^L\}$ of signals in $\cK_s$.
For each $i$, let $T_i \inc \{ 1, \ldots, N \}$
denote an index set of $t$ largest absolute entries of $\vD^* \vf^i$.
We write
$$
\rho < \|\vf^i - \vf^j \|_2 = \|\vD^* \vf^i - \vD^* \vf^j \|_2 
\le \|(\vD^* \vf^i)_{T_i} - (\vD^* \vf^j)_{T_j} \|_2
+ \| (\vD^* \vf^i)_{\ol{T_i}} \|_2 +  \| (\vD^* \vf^j)_{\ol{T_j}} \|_2.
$$
Invoking \cite[Theorem 2.5]{foucart2013}, we observe that
$$
\| (\vD^* \vf^i)_{\ol{T_i}} \|_2
\le \f{1}{2\sqrt{t}} \|\vD^* \vf^i \|_1 \le \f{\sqrt{ s}}{2 \sqrt{t}}
\|\vD^* \vf^i \|_2 = \f{\sqrt{ s}}{2 \sqrt{t}},
$$
and similarly for $j$ instead of $i$.
Thus, we obtain
$$
\rho <  \|(\vD^* \vf^i)_{T_i} - (\vD^* \vf^j)_{T_j} \|_2 + \sqrt{\f{s}{t}}
\le  \|(\vD^* \vf^i)_{T_i} - (\vD^* \vf^j)_{T_j} \|_2 + \f{\rho}{2},
\quad \mbox{i.e.,} \;
 \|(\vD^* \vf^i)_{T_i} - (\vD^* \vf^j)_{T_j} \|_2 > \f{\rho}{2}.
$$
Since we have uncovered a set of $L = \cP(\cK_s,\rho)$ points in $\Sigma_t^N \cap B_2^N$ that are $(\rho/2)$-separated,
the claimed inequality is proved. 
We conclude by recalling that $\cP(\Sigma_t^N \cap B_2^N, \rho/2)$
is bounded above by $\cN(\Sigma_t^N \cap B_2^N, \rho/4)$,
which is itself bounded above by $\dbinom{N}{t} \left( 1 + \df{2}{\rho/4} \right)^t$.
\epf

We can now turn our attention to proving the awaited theorem.

\bpf[Proof of Theorem \ref{ThmTesSphere}]
With $\cK_s = (\vD^*)^{-1}(\Sigma_s^{N,{\rm eff}}) \cap S^{n-1}$,
the conclusion holds when $m \ge C \eps^{-6} w(\cK_s)^2$
or when $m \ge C \eps^{-1} \ln (\cN(\cK_s,c \eps))$,
according to \cite[Theorem 1.5]{pv-embeddings}
or to \cite[Theorem 1.5]{bilyk2015random}, respectively.
It now suffices to call upon Lemma \ref{LemWNana}.
Note that the latter option yields better powers of $\eps^{-1}$
but less pleasant failure probability. 
\epf

\subsection{Further remarks}

We conclude this theoretical section by making two noteworthy comments on the sign product embedding property and the tessellation property in the dictionary case.

\brk
$\vD$-SPEP cannot hold for arbitrary dictionary $\vD$ if synthesis sparsity was replaced by effective synthesis sparsity. 
This is because the set of effectively $s$-synthesis-sparse signals can be the whole space $\bR^n$.
Indeed, let $\vf \in \bR^n$ that be written as $\vf = \vD \vu$ for some $\vu \in \bR^N$.
Let also pick an $(s-1)$-sparse vector $\vv \in \ker \vD$ --- there are tight frame for which this is possible, e.g. the concatenation of two orthogonal matrices.
For $\eps > 0$ small enough,
we have
$$
\f{\|\vv + \eps \vu \|_1}{\|\vv + \eps \vu\|_2}
\le \f{\|\vv\|_1 + \eps \|\vu\|_1}{\|\vv\|_2 - \eps \|\vu\|_2}
\le  \f{\sqrt{s-1} \|\vv\|_2 + \eps \|\vu\|_1}{\|\vv\|_2 - \eps \|\vu\|_2}
\le \sqrt{s},
$$  
so that the coefficient vector $\vv + \eps \vu$ is effectively $s$-sparse,
hence so is $(1/\eps)\vv + \vu$.
It follows that $\vf = \vD((1/\eps)\vv + \vu)$ is effectively $s$-synthesis sparse.
\erk

\brk
Theorem \ref{ThmSPEP} easily implies a tessellation result for $\vD(\Sigma_s^N) \cap S^{n-1}$, the ``synthesis sparse sphere''.
Precisely, under the assumptions of the theorem (with a change of the constant $C$), 
$\vD$-SPEP$(2s,\delta/2)$ holds.
Then,  one can derive
$$
[ \vg,\vh \in \vD(\Sigma_s) \cap S^{n-1} : \; \sgn(\vA \vg) = \sgn(\vA \vh) ]
\imp 
[ \|\vg - \vh\|_2 \le \delta ].
$$
To see this, with $\boldsymbol{\eps} := \sgn(\vA \vg) = \sgn(\vA \vh)$
and with $\vf := (\vg-\vh)/\|\vg-\vh\|_2 \in \vD(\Sigma_{2s}) \cap S^{n-1}$,
we have
$$
\left| \f{\sqrt{2/\pi}}{m} \langle \vA \vf , \boldsymbol{\eps}\rangle - \langle \vf, \vg \rangle \right| \le \f{\delta}{2},
\qquad
\left| \f{\sqrt{2/\pi}}{m} \langle  \vA \vf , \boldsymbol{\eps} \rangle - \langle \vf, \vh \rangle \right| \le \f{\delta}{2},
$$
so by the triangle inequality $|\langle \vf, \vg - \vh \rangle| \le \delta$,
i.e.,
$\|\vg -\vh\|_2 \le \delta$, as announced.
\erk

\section*{Acknowledgment}
The authors would like to thank the AIM SQuaRE program that funded and hosted our initial collaboration. In addition,
Baraniuk is partially supported by NSF grant number CCF-1527501, ARO grant number W911NF-15-1-0316, AFOSR grant number FA9550-14-1-0088,
Needell by an Alfred P. Sloan Fellowship and NSF Career grant number 1348721,
Plan by NSERC grant number 22R23068, and
Wootters by NSF Postdoctoral Research Fellowship grant number 1400558.

\bibliographystyle{myalpha}
\bibliography{Dict1BitBib}

\end{document}